\documentclass[11pt]{article}
\usepackage{authblk}
\usepackage{graphicx}
\usepackage{xcolor}
\usepackage{natbib}
\usepackage[margin=2cm]{geometry}

\title{\textbf{Optical SETI at ESO in the 2040s\footnote{This is an 
extended version of a white paper submitted in response to the 
ESO Expanding Horizons initiative.}}}

\author[1]{Valenitn D. Ivanov}
\affil[1]{European Southern Observatory, Germany}

\date{December 2025}

\begin{document}

\maketitle
\thispagestyle{empty}

\abstract{
The searches for other life and for intelligence are fundamental problems
that science faces today. Most searches so far have been focused
on radio, but optical laser communication is an alternative, well suited 
for a ground-based observatory. A project to search for artificial laser 
communications with the current and future extreme multiplexity 
spectroscopic facilities that ESO may develop by the 2040s is outlined. 
The monochromatic light is a clearly identifiable technosignature. The
enormous corollary outreach potential of this initiative is underlined.
}

\section{Signatures of life}
Finding other life in the Universe is one of the most fascinating tasks
of science. Even the details of the origin of life on our own planet 
are unclear and therefore a second example is needed to better understand 
this process. The basic building blocks of life -- at least of life 
similar to ours -- are commonly found across the cosmos 
\citep{2021isms.confERC06M}, but so far we have not securely identified 
life elsewhere \citep[notwithstanding a few questioned 
claims:][]{1996Sci...273..924M,2021NatAs...5..655G,2023ApJ...956L..13M}, 
including other planets in the Solar System where in situ exploration is 
possible.

However, the search space is widening. So far we know of over 6000
exoplanets and the orbits of $\sim$70 of them reside in the so-called 
habitable zone \citep[HZ;][]{2025arXiv250114054B} -- a loosely defined 
region where the equilibrium temperature T$_{eq}$ of the exoplanet allows 
the existence of liquid water. HZ is a somewhat misleading term, because 
T$_{eq}$ also depends on other parameters, such as the planet mass, if 
there is greenhouse effect, orbital eccentricity, if the host star 
exhibits high-energy flares that could photoevaporate the planetary 
atmosphere, etc.

The are two avenues to search for life:

{\bf(I) Biosignatures} -- spectral features in exoatmosphere from
various volatile molecules that can originate from life (O$_2$, O$_3$,
CH$_4$, N$_2$O, CH$_3$Cl, etc.), or that indicate reflection from
biological material (e.g., vegetation red edge); the seasonal
variation
of these signatures can also suggest the presence of life. However,
many of these species can also originate from geological processes,
leaving the possibility for false positives. A safer strategy is to
look for a complex of them. Biosignatures from our 
own life in the Earth's atmosphere are a major obstacle, suggesting 
that these searches are likely to be more successful with space-based 
facilities -- a path suggested early on by \citet{1992ESASP.354...81B} 
and recently adopted by the Habitable Worlds Observatory 
\citep{2020arXiv200106683G}.

{\bf(II) Technosigantures} -- indications of technology (either
because of their nature or because of their information content) that
cannot be produced by natural processes. The detection of radio signals 
was the first to be seriously considered and attempted 
\citep{1959Natur.184..844C,1964SvA.....8..217K}, together with thermal 
residual emission from Dyson spheres 
\citep{1960Sci...131.1667D,2000AcAau..46..655T,2009ApJ...698.2075C}
and later -- transits from artificial megastructures such as Dyson 
swarms \citep{2016ApJ...816...17W}, among other examples of searches.

A common problem of radio and thermal technosignatures is that they 
rely on wasted radiation from the extraterrestrial civilizations. We 
already see in our own example the emerging tendency to reduce waste 
and to become more efficient. \citet{2020A&A...639A..94I} considered 
``quiet'' (and energy-savvy) advanced civilizations that may co-exist 
with us, yet they would remain invisible to our searches. The authors
conclude that if this evolutionary trend toward more rational (and 
more economical) existence is common among civilizations, then our 
best detection opportunities are with the ones at technological 
stages of development close to our own, that also happened to be 
located nearby, or with the significantly more advanced civilizations 
that would set up beacons, purposely designed and optimized to be 
detectable by younger ``cousins'' like us 
\citep{2010AsBio..10..475B,2010AsBio..10..491B}.

There was an early suggestion to use an optical communication channel, 
that does not rely on wasted energy. Carl Friedrich Gauss is said to 
have proposed in 1820 illuminated drawings of regular shapes in the 
deserts, intended to be visible from other planets in the Solar System. 
The optical signals became relevant for interstellar distances after 
\citet{1961Natur.190..205S} suggested using lasers for communication. 
\citet{2004ApJ...613.1270H} carried out a search,
\citet{2014SPIE.9147E..4KM} moved it to the infrared, minimizing the 
interstellar dust obscuration effects. \citet{2002PASP..114..416R} 
begun a massive archival search for narrow unresolved emission lines 
in the high-resolution spectra used for radial velocity planet 
searches. The increase of resolution suppresses the contribution of 
the continuous spectrum in the spectral resolution element where the 
lased light falls.

{\bf Laser communications, unlike radio and thermal waste, are highly 
directional, so the senders must intentionally attempt to communicate 
with us, or at least they must have included the Solar System in their 
list of promising targets.}

\section{The Solar System as a target of directional signaling}

\citet{2001ARA&A..39..511T} points out that SETI (Search for 
Extraterrestrial Intelligence) programs are easier to get approved and 
carried out, and perhaps even more likely to succeed, if they are 
performed in the course of regular astronomical observations, as 
opposed to dedicated SETI campaigns. Given the vast parametric space 
that SETI needs to cover and the poorly constrained properties of 
extraterrestrial communication, the astronomical surveys stand out as
the most obvious types of programs to look for synergies with, due to
of their capability to capture diverse signals: potentially from wide 
areas on the sky and in broad wavelength ranges.

One can arrive at the idea of using the established channel of the 
astronomical observations to signal other civilizations from a game 
theory concept -- that of communication without 
messaging\footnote{{\it The Strategy of Conflict}, Thomas Schelling
(1960).}. Despite its complicated name, this is something most people 
apply in everyday life, without even realizing it, when they ask 
themselves the question what is expected of them by others. Card games 
where one pair of players opposes another pair are a typical example 
of cooperation without message exchange where the rules and goals help 
players coordinate. In daily life we have the laws, social, ethical, 
and religious norms, and the common sense as analogs of the game rules.

The only assumption necessary in the context of SETI is that the other
civilizations are interested in the Cosmos to the extent of exploring 
it by means of astronomical observations; the homogeneity of the
Universe ``unifies'' the communication channels of distant
civilizations. For example, for a radioastronomer -- human or alien --
the radio noise in the Milky Way will be the lowest at 0.3-30\,cm
\citep[][fig.\,1]{1964SvA.....8..217K}.

Following similar reasoning \citet{1991LNP...390..254F,2016AsBio..16..259H} 
arrived at the idea that astronomical observations can reveal the 
Earth's habitability via transmission spectroscopy observable by 
extraterrestrial astronomer located in out ecliptic plane 
\citep[e.g.,][]{2002ApJ...568..377C}. This is possible via emission 
spectroscopy \citep[e.g.,][]{2005Natur.434..740D} for observers away 
from the ecliptic plane as well, but it is a more challenging 
observation for cool exoplanets in the HZ.

\section{Technosignature search opportunities at ESO with surveys from 
today to the 2040s}

ESO has a long and successful tradition in astronomical surveys, and 
is entering a new era of high-multiplexity spectroscopic surveys with 
the 4-metre Multi-Object Spectroscopic Telescope 
\citep[4MOST, to begin operations in 2026;][]{2019Msngr.175....3D}.
Furthermore, a new 10-m class Wide-field Spectroscopic Telescope 
\citep[WST;][]{2024arXiv240305398M} with an extremely high multiplexity 
spectrograph ($\sim$30,000 optical fibers) is being proposed for the 
2040s. 
4MOST and WST are especially well suited for optical SETI, searching for 
laser communications that will appear as unresolved emission lines in 
spectra of stars that do not show natural features like those. The 
program discussed here  relies on the synergy with other surveys -- it 
can be carried out in parallel with any stellar survey.

\section{Timeline, facilities and requirements}

The search for laser communications requires:
\vspace{1mm}

{\bf (I) High multiplexity} to maximize the number of observed stars.
\vspace{1mm}

{\bf (II) High cadence} to increase the signal-to-noise ratio (SNR), if 
the senders have adopted an energy-efficient strategy that would minimize 
the energy expenditure -- the host star emits continuously, while the 
laser pulses are expected to be brief, so shorter integrations improve 
detection significance.
\vspace{1mm}

{\bf (III) High spectral resolution}, again to increase the SNR, because 
the laser emits monochromatic light superimposed ob the continuous stellar 
spectrum, so the monochromatic laser pulses will be confined to a single 
spectral resolution element, and the narrower this element is, the higher 
the SNR.
\vspace{1mm}

{\bf (IV) High processing speed} to allow rapid discovery and scheduling 
of follow-up observations.
\vspace{1mm}

Meeting the scientific goals requires more than high-cadence spectroscopy: 
a data flow system capable of processing data, detecting signals and 
interpreting them in real time is needed. The delays can undermine the 
follow-up. No existing facility offers these capabilities and the planned 
WST will be {\bf a giant step for the optical SETI}.

\section{Outreach potential}

A long-term future SETI project at ESO can become an excellent vehicle 
for education, for explaining to the public the most basic scientific 
concepts like the scientific method, for introducing the newest 
discoveries -- and reaching far more people than usual channels
\citep{2021arXiv211214702I,2023arXiv231215797I,2023arXiv231216254I}.

The interstellar comet 3I/ATLAS helped to demonstrate the significant
public interest in the search for life and its implications. This 
object attracted an enormous attention, because of the possibility -- 
however remote -- that it may be a piece of alien technology 
\citep[e.g.,][]{2025arXiv250712213H}. An opportunity to study a comet 
originating from a different planetary system can be a treasure trove 
for new discoveries, but the public reaction is important on its own 
because it demonstrated how much the imagination and curiosity of 
people can be aroused by the unanswered questions of the Universe.

\section{Summary and conclusions}

A SETI program at ESO in the next decade will allow addressing a 
question of both scientific and public interest -- if intelligent
life exists elsewhere in the Universe. We consider here a program 
that can be executed in parallel with the new large scale 
spectroscopic surveys that will be carried out in the next coming 
years, and with the future WST facility that has been proposed as 
ESO's next big telescope project.

\section*{Acknowledgements}
The author is grateful to Claudio C\'aceres Acevedo, Dante Minniti,
and Juan Carlos Beam\'in for the fruitful discussions.

{\small
\bibliographystyle{abbrvnat}
\setlength{\bibsep}{-2pt}
\bibliography{biblio}
}

\end{document}